\documentstyle[osa,aps,prb,epsfig]{revtex}
      \begin{document}

      \twocolumn[\hsize\textwidth\columnwidth\hsize\csname
      @twocolumnfalse\endcsname

      \title{$^{\ast\ast}$Asymmetry in fatigue and recovery
      in ferroelectric Pb(Zr,Ti)O$\bf{_3}$ thin-film capacitors}

      \author{B. G. Chae, C. H. Park$^{\dagger}$, Y. S. Yang, and M. S. Jang}
      \address{Research Center for Dielectric and Advanced Matter Physics,\\
      Pusan National University, Pusan 609-735, Korea}

      \maketitle{}

      \begin{abstract}
      We investigate the fatigue and refreshment by dc-electrical field
      of the electrical properties of Pt/Pb(Ti,Zr)O$_3$/Pt ferroelectric capacitors.
      We find an asymmetry in the refreshment, that is, the fatigued state
      can be refreshed by application of negative high dc-voltage to the top electrode,
      but no refreshment is measured by positive dc-voltage application.
      We also find that the fatigue can be prevented by driving the capacitor
      asymmetrically.
      \\
      \\
      \end{abstract}
      ]

      Perovskite ferroelectrics are widely investigated
      since the perovskite compounds can be applied
      in developing low energy-consuming and high-speed
      semiconductor memories.\cite{1,2,3,4,5}
      A problem to be resolved for its application is
      fatigue.\cite{6,7}
      It has been recently extensively investigated.\cite{8,9,10,11}
      The fatigue can be prevented by the use of metal-oxide
      electrodes\cite{12,13} or layered perovskite ferroelectrics,\cite{14,15}
      however, the switchable polarization in these is small and
      the control of the electrical properties such as leakage current
      is not easy.
      Many kinds of explanations for the fatigue have been
      suggested such as large-scale defect migration,\cite{8,16} domain pinning
      by defects or grain boundary,\cite{9,17}
      and the screening at electrode interface;\cite{10,18}
      however the fatigue mechanism is not yet clearly understood.

      An important property of the phenomena is that the fatigued
      polarization is refreshed by applying dc-electrical field,\cite{7,19}
      heat treatment,\cite{20} or UV-light illumination.\cite{9}
      In finding the microscopic origin of the fatigue phenomena,
      the refreshment should be explained.
      However, the behavior of the refreshment is not frequently investigated.
      In this letter, we investigate the refreshment behaviors during applying dc
      electrical fields in the fatigued Pt/PZT/Pt capacitor.

      The Pb(Zr$_{0.52}$Ti$_{0.48}$)O$_3$ films were grown on
      Pt/Ti/SiO$_2$/Si substrates by the sol-gel method.\cite{21}
      The ferroelectric films were deposited by a multilayer spin-coating
      technique\cite{21} and crystallized in air at 650$\rm{{}^oC}$
      for 30 min. The details were described in our recent articles.\cite{7,22}
      The perovskite ferroelectric films are measured to
      be oriented mainly along the [111] direction from X-ray diffraction patterns,
      and to be polycrystalline by scanning electron microscopy.
      The thickness of the PZT films is measured to be 300 nm.
      We fabricate the Pt top electrode with the area of 100$\times$100
      $\mu$$\rm{m^2}$ by sputtering through the shadow mask
      at a temperature of around 500$\rm{{}^oC}$.

      The polarizations of the PZT capacitor were repetitively
      switched by the square electrical pulses generated by a function generator
      connected to the RT66A ferroelectric tester.
      The measurement of the polarization was carried out by use of
      triangular shape of electrical pulse, whose peak voltage is
      $\pm$ 5 V which corresponds to the electrical field intensity of $\pm$ 167 kV/cm.
      For the refresh experiments, the bottom electrode of the thin film capacitor is grounded
      and the various voltages from 0 to $\pm$ 15 V were applied to the top
      electrode of the fatigued films.

      Figures 1(a) and 1(b) describes the fatigue in the switchable
      polarization (P*-P$^\wedge$) of a Pt/PZT/Pt capacitor.
      As shown in Fig. 1(a), as the peak voltage of the switching pulses
      increases from 3 V to 7 V,
      the switchable polarization is measured to fatigue faster.
      The fatigue starts around after 10$^{4}$ cyclings of switching,
      nearly independently to the peak voltages ranging from 3 V to 7 V.
      We also examined the fatigue as the variation
      of the frequency of the switching pulses.
      As described in Fig. 1(b), the development of the fatigue is dependent
      mainly on the number of the polarization switching.
      These indicate that the fatigue in the present Pt/PZT/Pt
      capacitor progresses mostly during the switching.\cite{6}

      We measured the change of the remanent polarization
      during the application of the dc-field to the fatigued ferroelectric
      capacitors, as described in Fig. 2.
      When we applied the dc voltage of -10 V
      to the top electrode of the fatigued Pt/PZT/Pt capacitors,
      the remanent polarization is measured to be recovered very fast,
      up to 70 \% of the initial value just after 1 s
      and nearly completely recovered after 100 s.
      By the application of the weaker electrical field,
      it is less rapidly recovered.
      After the application of the dc field for 100 s,
      the measured remanent polarization versus the applied voltage
      are shown in Fig. 3.
      We find that only if the dc voltage is larger
      than the coercive voltage, the polarization can be recovered eventually
      nearly to the initial value.

      A surprising result is that the fatigued polarization is $little$ refreshed
      by the application of $positive$ dc voltage.
      In Fig. 2, the change of the polarization is described
      by the application of $\pm$10 V.
      Only application of $negative$ dc-voltage makes the polarization refreshed.
      The asymmetricity of the recovery should be related to
      the properties of the thin-film Pt/PZT/Pt capacitor, which will be discussed
      later.

      We examined the re-fatigue of the recovered state.
      The recovered state is found to be still resistive to fatigue,
      however, the switchable polarization is a little more
      rapidly fatigued, ten times faster than the initial state, indicating
      that the recovered state is somewhat different from the initial as-grown state.
      But as the fatigues and refreshments are repeated twice and further,
      the behaviors of the further refatigues are found to be similar to that
      of the first refatigue.
      We examined the recovery by the heat-treatment above Curie temperature.
      The re-fatigue behavior is similar to the above case.
      This indicates that there should be some redistribution of ions and defects
      during the fatigue that are not completely recovered by the dc field for a short
      time.

      The asymmetricity of the recovery indicates that there should be
      a similar asymmetric behavior of the fatigue.
      We examined the latter asymmetricity by applying asymmetrical
      sequence of electrical pulses to the capacitors in switching
      their polarizations, as described in the inset of Fig. 4,
      where $V_+^p$ is set to be different from $V_-^p$.
      The fatigue behaviors are described in Fig. 4, and Fig. 5 describes
      the measured switchable polarization
      after 4$\times$$\rm{10^9}$ cycles of switchings versus the asymmetricity $A$.
      Here, we define the asymmetricity $A$ of the asymmetric switching pulses
      by $\Delta$V/($V_+^p+V_-^p$) where
      $\Delta V$ is $V_+^p$$-$$V_-^p$.
      We fix the sum of $V_-^p$ and $V_-^p$ to be 10 V
      and the values of $V_+^p$ and $V_-^p$ were set to be at least greater than 2 V,
      which is larger than the coercive voltage of 1.5 V, so that the polarization
      can be switched.
      The negative $A$ means that $V_+^p$ is smaller
      than $V_-^p$.

      It is remarkable that in the case that $A$ is less than -0.3
      the switchable polarization is found to be little fatigued,
      even after 4$\times$$\rm{10^9}$ cycles of polarization switching.
      In the inset of Fig. 5, the measured hysteresis loops of the polarizations
      before and after the 4$\times$$\rm{10^9}$ switching cycles
      are compared in the cases of $A$ = -0.4.
      The hysteresis loop is only slightly right-shifted
      with an increase of the coercive field from the initial 52 kV/cm to 70 kV/cm.
      In the case of the positive asymmetricity $A$, the polarization
      fatigues more rapidly compared to the normal case of zero $A$.

      We would note that recently a similar asymmetric behavior is reported;\cite{23}
      the direct observation of the pinned domain using atomic force microscopy (AFM) indicated
      that the pinned domain associated with the fatigue has
      a preferential orientation of polarization.

      We suggest that these asymmetric behaviors in the fatigue and refresh
      are related to the asymmetric distribution of fatigue centers
      in the PZT thin film.
      Many experimental evidences indicate that the defects such as oxygen-vacancies
      are a source of the fatigue phenomena.\cite{8,9,17,18,24,25}
      It is well known that the concentration of defects such as oxygen deficiency
      and lead-vacancies in the ferroelectric films
      are more significant around the bottom electrode than around
      the top electrode,\cite{25,26,27}
      therefore, it is suggested that fatigue centers should be developed dominantly
      around the bottom electrode.

      Electron paramagnetic resonance data\cite{28}
      indicated that the charge trappings at defects are accompanied with fatigue.
      Theoretical calculations\cite{29}
      indicated that oxygen vacancy induce tail-to-tail polarization
      around itself, and that the hole trap or electron ionization
      at oxygen vacancies can enhance the vacancy-induced polarization
      and indicated that their cooperative forces through their accumulation
      during the repetitive switching of the polarization
      can lead to strong domain pinning, becoming the fatigue center.\cite{17,30}
      Since the oxygen vacancies are accumulated more around the bottom electrodes,
      the polarization direction of the pinned domain from these defect
      should preferably be from the bottom toward the top electrode.

      In the refreshment by the application of the negative dc-voltage
      to top electrode, the positive carriers captured at the O vacancies
      around the interface of the bottom electrodes can migrate
      toward the top electrode, leading to the refreshment.
      Only if the polarization is inverted by a dc field larger than the coercive field,
      the region of the antiphase polarizations from the oxygen vacancies
      around interfaces,
      which leads to the capture of hole at the vacancies,\cite{29}
      is eliminated and the hole trapping at the defects should be reduced.
      However, by the positive voltage application, the migration of the captured
      holes can be prevented by the Schottky barrier between ferroelectrics
      and the bottom electrode.
      The PZT films are usully slightly $p$ type doped due to the Pb-vacancies,
      which makes a Schottky barrier preventing hole diffusion
      to the bottom-electrode.
      Therefore only the negative voltage application can refresh the fatigue.
      For a similar reason, the negative $A$ asymmetric
      switching pulses can prevent the capture of holes at oxygen
      vacancies around the bottom electrode, which leads to the present fatigue-resistive
      behavior.
      On the other hand, the suggested accumulation\cite{17,30} of defects
      during fatigue can give an explanation to the re-fatigue behavior
      after the refreshment that is mentioned above.
      The accumulation of defects may not
      be eliminated by the application of short-time dc-electrical pulses,
      therefore the refatigue after the refreshment can proceed more rapidly,
      since the refatigue can be developed
      just by the capture of carriers at these accumulated defects.

      It is recently reported that the ferroelectric
      capacitor by the use of $n$-type semiconductor at the top
      electrodes is more easily fatigued,\cite{27}
      suggesting that electron captures at defects or interfaces
      are more effective in fatigue.
      We would suggest that the experimental data should be explained in the respect
      of the defect formations during the thin-film process
      rather than charge trapping,
      since the ferroelectrics/doped-semiconductor does not
      make the conventional p-n junction
      and the current density of injected carrier is not high, compared to the
      available leakage current.

      This work was supported by the Korea Science and Engineering Foundation
      (KOSEF) through the Research Center for Dielectric and Advanced Matter Physics
      (RCDAMP) at Pusan National University.

      \begin{figure}
      \vspace{2.0cm}
      \centerline{\epsfysize=11cm\epsfxsize=8.5cm\epsfbox{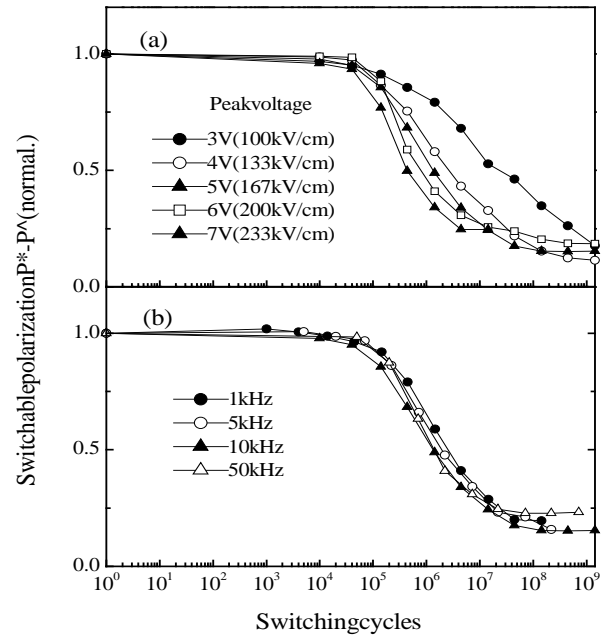}}
      \vspace{-1.0cm}
      \caption{Changes of the switchable polarizations (P*-P$^\wedge$)
      of Pt/PZT/Pt capacitor during the repetitive switching are described,
      where we varied (a) the peak voltage and (b) the frequency of the switching
      pulses.
      The values are divided by the initial value of the polarization
      which is measured to be 32.7 $\mu$C/$\rm{cm^2}$.}
      \end{figure}

      \begin{figure}
      \vspace{-1.0cm}
      \centerline{\epsfysize=11cm\epsfxsize=8.5cm\epsfbox{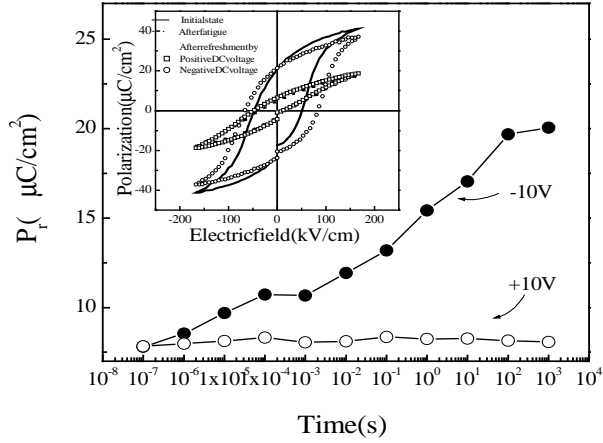}}
      \vspace{-2.0cm}
      \caption{The change of the remanent polarization during the refreshment
      by the application of the dc electric field to the fatigued films.
      (solid circles for the application of +10 V to the top electrode
      and open circles for the application of -10 V).
      The inset desribes the changes of the hysteresis loops
      before and after fatigue and after the refreshments.}
      \end{figure}

      \begin{figure}
      \vspace{-1.0cm}
      \centerline{\epsfysize=11cm\epsfxsize=8.5cm\epsfbox{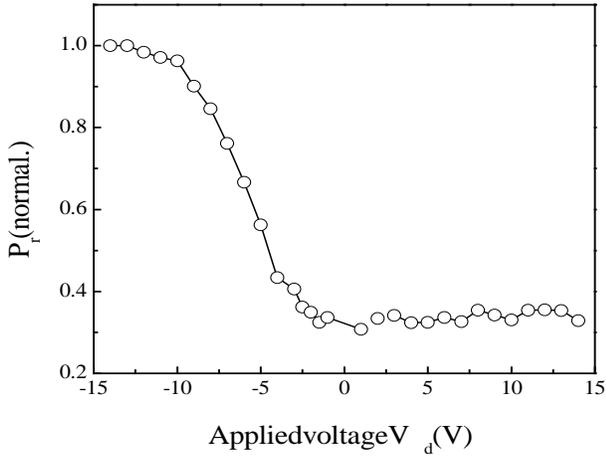}}
      \vspace{-2.0cm}
      \caption{The measured remanent polarizations after the refreshments
      for 100 s vs the magnitute of the applied dc voltage.
      The values are divided by the initial value of the polarization before fatigue.}
      \end{figure}

      \begin{figure}
      \vspace{-1.0cm}
      \centerline{\epsfysize=11cm\epsfxsize=8.5cm\epsfbox{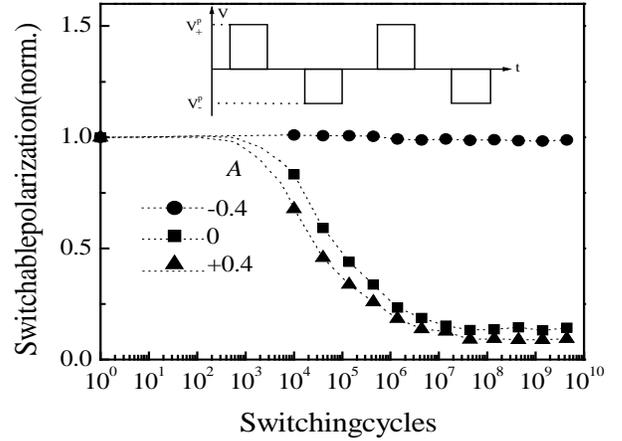}}
      \vspace{-2.0cm}
      \caption{Fatigue behaviors by the use of the asymmetric switching driving pulses
      which are described in the inset and whose frequency is 10 kHz.
      The asymmetricity is described by $A$ that is ($V_+^p-V_-^p$)/($V_+^p+V_-^p$).
      The values of the switchable polarizations are divided by the initial
      value.}
      \end{figure}

      \begin{figure}
      \vspace{-2.0cm}
      \centerline{\epsfysize=11cm\epsfxsize=8.5cm\epsfbox{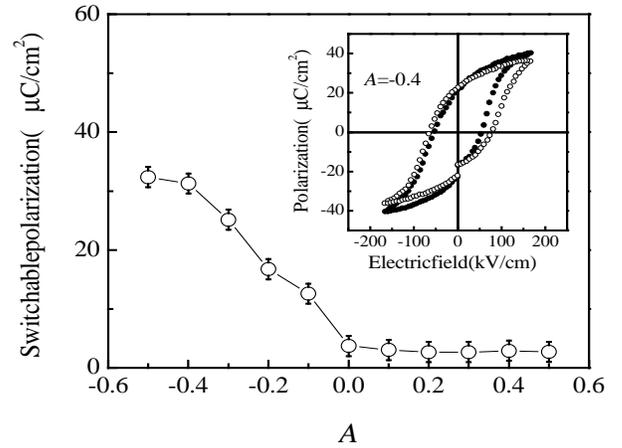}}
      \vspace{-2.0cm}
      \caption{The switchable polarizations after 4$\times$$\rm{10^9}$ cycles switching
      vs the asymmetricity of the switching pulses.
      The inset shows the hysteresis loops before and after the switchings
      when the asymmetricity $A$ is -0.4.}
      \end{figure}

      \end{document}